\documentclass[acmsmall]{acmart}
\AtBeginDocument{%
  }

\acmConference[CSCW '26]{ACM CSCW 2026}{<conference dates>}{Salt Lake City, Utah, USA}
\usepackage{color}

\usepackage{tabularray}
  \usepackage{changepage}
\newcommand{\qj}{\textcolor{black}}
\newcommand{\qjnew}{\textcolor{black}}
\newcommand{\abnew}{\textcolor{black}}

\newcommand{\abrev}{\textcolor{black}}
\newcommand{\final}{\textcolor{black}}

\usepackage{booktabs}
\usepackage{tabularx}

\begin{document}

\title{Understanding \abrev{Newcomer Persistence} in Social VR: A Case Study of VRChat}

\author{Qijia Chen}
\email{qijia.chen@helsinki.fi}
\orcid{0000-0003-1038-5461}
\affiliation{%
  \institution{University of Helsinki}
  \streetaddress{Yliopistonkatu 4}
  \city{Helsinki}
  \postcode{00100}
   \country{Finland}
}

\author{Andrea Bellucci}
\email{abellucc@inf.uc3m.es}
\orcid{0000-0003-4035-5271}
\affiliation{%
  \institution{Universidad Carlos III de Madrid }
  \streetaddress{Avenida de la Universidad, 30}
  \city{Leganés}
  \postcode{28911}
  \country{Spain}
  }

\author{Giulio Jacucci}
\email{giulio.jacucci@helsinki.fi}
\orcid{0000-0002-9185-7928}
\affiliation{
  \institution{University of Helsinki}
  \streetaddress{Yliopistonkatu 4}
  \city{Helsinki}
  \country{Finland}
  }

\begin{abstract}
Newcomers are crucial for the growth of online communities, yet \abrev{their successful integration into these spaces requires overcoming significant initial hurdles}. Social Virtual Reality (VR) platforms are novel avenues that offer unprecedented online interaction experiences. Unlike well-studied two-dimensional online environments, the \abrev{pathways to successful} newcomer integration in online VR spaces are underexplored. Our research addresses this gap by examining the \abrev{strategies used by newcomers to navigate early challenges} in social VR and how they adapt. \abrev{By focusing on active participants (ranging from newcomers currently navigating these hurdles to veterans who have successfully integrated) we isolate the specific strategies necessary for retention.} We interviewed 24 active social VR users and conducted a reflexive thematic analysis. \abrev{While participants identified barriers such as} unfamiliar user interfaces, social norms, and overwhelming sensory input, \abrev{our analysis reveals the adaptation strategies required to overcome them}. Our findings expand on understanding newcomer \abrev{persistence} beyond traditional 2D environments, emphasizing \abrev{how social dynamics influence the management of} VR-specific issues like VR sickness during onboarding. Additionally, \abrev{we highlight how successful newcomers overcome} the lack of clear objectives in social VR \abrev{by proactively constructing social meaning}. We propose design suggestions to \abrev{scaffold these successful integration pathways}.
\end{abstract}



\maketitle
\section{Introduction}

Social Virtual Reality (VR) platforms like VRChat offer immersive three-dimensional (3D) environments where users interact through avatars and spatial audio~\cite{moustafa2018longitudinal}, enabling novel forms of social engagement that closely resemble in-person encounters~\cite{maloney2020talking}. These environments support diverse forms of engagement, including social, recreational, and intimate interactions~\cite{Chen05, Gunkel66, freeman2021hugging, Yin2023}. As VR hardware becomes more accessible, these platforms are growing rapidly and attracting attention from both users and researchers; \abrev{however, they remain volatile environments for new users, often plagued by issues of safety, harassment, and inadequate moderation~\cite{weerasinghe2025beyond, freeman2022disturbing}}. Recognizing the uniqueness of social VR, prior studies have examined various aspects of the spaces, including novel communication dynamics~\cite{chen2025understanding, maloney2020talking}, new forms of harassment~\cite{freeman2022disturbing, Blackwell22}, and platform design strategies~\cite{mcveigh2019shaping, 9258343}. However, \abrev{while the community understand the potential of these spaces}, there remains a gap in understanding \abrev{how newcomers successfully navigate the transition from outsider to insider}. 

The emergence of social VR has introduced a shift in digital interaction, pushing the boundaries of virtual communities into immersive 3D spaces~\cite{sykownik2021most}. Newcomers are vital to sustaining online communities, bringing new perspectives and replenishing user bases~\cite{benkler2015peer, forte2013defining, Kiene356}. Yet, \abrev{the threshold for entry is high.} Newcomers are particularly vulnerable to early setbacks and are more likely than established users to disengage after negative experiences~\cite{kraut2012building, Singh2012Newcomer, Morgan2018Evaluating}. 
\abrev{While researchers have identified key obstacles that drive users away, such as technical friction~\cite{maloney2020talking, stanney2020virtual} or toxic interactions~\cite{blackwell2019harassment, weerasinghe2025beyond}, far less is known about the adaptation strategies that allow newcomers to persist in social VR environments.}

Previous extensive research has focused on newcomer integration in online spaces~\cite{Morgan2018Evaluating, Steinmacher2015Social,steinmacher2013newcomers}. Nonetheless, much of this research has focused on 2D environments, predominantly within collaborative online spaces such as open-source programming projects and Wikipedia~\cite{Jensen2011Joining, Morgan2018Evaluating} \abrev{ or structured Massively Multiplayer Online Games (MMOGs) where clear objectives guide the user \cite{ducheneaut2005more}. Earlier accounts of text-based multi-user dungeons  documented how newcomers faced steep learning curves, ambiguous norms, and harassment and how such issues were often mitigated through community rituals and governance structures~\cite{dibbell1994rape, reid1994cultural}. While these studies have shaped our understanding of online integration, social VR introduces a new layer of complexity, where embodiment, sensory immersion, and emergent social conventions significantly reshape the newcomer experience~\cite{freeman2021hugging, maloney2020talking, schulenberg2023towards}.}

Given the novelty of social VR, it is important to understand \abrev{not just the challenges newcomers face, but the specific strategies they construct to integrate into these online spaces.}. Our research \abrev{examines the adaptation strategies} newcomers encounter when entering social VR, and the mechanisms through which they acclimate and integrate into these settings. \abrev{By focusing on the lived experiences of active users, we isolate the persistence factors that are essential for retention.} Exploring these \abrev{successful integration pathways of persisting newcomers} potentially provides insights that can guide the design and development of VR platforms, ensuring they are more inclusive, supportive, and engaging.

In the research, we interviewed 24 \abrev{active} users from VRChat\footnote{https://hello.vrchat.com/}, \abrev{ranging from newcomers currently navigating these hurdles to veterans who have successfully integrated}, and employed reflexive thematic analysis to examine the data. Our findings contribute to CSCW and the broader HCI community in several ways:

(1) We highlight \abrev{how newcomers navigate challenges} of social VR, such as \abrev{negotiating} unfamiliar user interfaces, understanding distinct social norms specific to VR, and \abrev{socially} managing overwhelming sensory input \abrev{(including VR sickness)}. Our research informs the design of VR onboarding, which traditionally emphasizes operational aspects like controller use in offline settings. 
(2) We underscore \abrev{the strategies newcomers use to overcome the lack of} clear objectives or narratives \abrev{inherent to open-world social VR}. \abrev{We show how the} absence of structure \abrev{requires newcomers to proactively construct social meaning to avoid} disorientation and decreased satisfaction. 
(3) We uncover \abrev{the specific integration pathways} newcomers \abrev{traverse} to familiarize themselves with online VR platforms, shedding light on \abrev{active} aspects of the integration process that have been previously overlooked. Prior research has mainly focused on designing \abrev{platform-side} strategies to support user integration, whereas our study explores user-driven integration processes, providing a more comprehensive understanding.
(4) Based on our findings, \abrev{we contribute a framework of newcomer persistence, identifying three distinct stages of adaptation: Acclimatization, Acculturation, and Embedding. This framework extends existing research by mapping how active users traverse specific friction points.} We propose design suggestions to \abrev{scaffold these successful pathways}, facilitate navigation of new social and technical environments, and enhance support systems to better integrate and retain users in social VR spaces.

\section{Related Works}

\subsection{Social VR as a Novel Place}
Social VR platforms create immersive, shared environments where people meet and interact in real time, emphasizing open-ended exploration ~\cite{mcveigh2019shaping, mcveigh2018s}. Platforms such as \emph{VRChat}, \emph{Meta Horizon Worlds}\footnote{\url{https://horizon.meta.com/}}, and \emph{Sansar}\footnote{\url{https://www.sansar.com/events}} function as open virtual worlds~\cite{kolesnichenko2019understanding, maloney2021social}, where users shape their experiences through user-generated content, embodied interaction, and social exploration \cite{chen2026social,chen2026usage,chen2025mirror}. 
This stands in contrast to immersive multiplayer online games, such as \emph{Gorilla Tag}\footnote{\final{\url{https://en.wikipedia.org/wiki/Gorilla_Tag}}}, which, while supporting real-time interaction, center on predefined game mechanics and competitive objectives.
Prior work often compares social VR to earlier virtual worlds like \emph{Second Life}\footnote{\url{https://secondlife.com/}} and \emph{Active Worlds}\footnote{\url{https://www.activeworlds.com/}}~\cite{barreda2022psychological, krell2023corporeal, maloney2021social}. While these predecessors also emphasized user-driven experiences, social VR introduces deeper immersion and embodiment through VR technology~\cite{barreda2022psychological}. Researchers have even described social VR as a precursor to the Metaverse, highlighting its lack of predefined goals, emphasis on personalization, and real-time embodied interaction~\cite{ortiz2022risks}.%
\footnote{While \emph{Rec Room} is sometimes categorized as social VR, its emphasis on competitive gameplay and leveling systems aligns it more closely with \qj{immersive multiplayer online games}. We therefore consider it an atypical, hybrid example.}

These environments leverage the affordances of VR to enable \abrev{high-fidelity non-verbal communication that goes beyond standard avatar movement~\cite{maloney2020talking}. Users do not merely control a character; they inhabit it, using head tracking to signal attention (gaze) and hand controllers to perform gestures like waving or pointing~\cite{li2019measuring, moustafa2018longitudinal}. This allows for nuanced interactions unavailable in 2D spaces, such as ``head pats'' (a common form of affection in VRChat) or high-fives that rely on synchronous timing~\cite{maloney2020talking}. Consequently, social VR is often perceived as more natural and emotionally engaging than screen-based platforms~\cite{maloney2020falling,chen2024understanding}.}

However, \abrev{this high social bandwidth introduces unique friction points that are particularly heightened for newcomers. Unlike text-based platforms where users can compose responses asynchronously, Social VR demands immediate, embodied improvisation. Research has consistently shown that many users experience pronounced social anxiety when engaging in VR environments (e.g., \cite{Chen05}). } Norms are emergent and often implicit. \abrev{For instance, ``personal space'' in VR is not a hard-coded game mechanic but a social expectation~\cite{yee2007unbearable}. A newcomer might accidentally ``clip'' through another user's avatar, violating unwritten rules of proxemics and causing social awkwardness that feels visceral due to the immersive nature of VR~\cite{williamson2021proxemics}. Furthermore, behaviors acceptable in gaming contexts (e.g., erratic movements) may be interpreted as aggression in social contexts~\cite{Blackwell22}, creating a steep learning curve for those transitioning from MMOGs who must relearn how to inhabit a virtual body in a socially acceptable way.}

\abrev{Beyond these interactional complexities, the broader landscape of Social VR is characterized by significant safety and moderation concerns. Prior work has highlighted the prevalence of harassment and the critical need for safety tools in immersive environments~\cite{weerasinghe2025beyond,chen2024people}. Research indicates that the embodied nature of VR can intensify negative social interactions, requiring distinct moderation strategies compared to 2D platforms \cite{freeman2022disturbing,chen2025democratic,chen2025investigating}, a challenge that is particularly acute regarding children's use of social VR~\cite{fiani2024exploring}. 
}

\abrev{However, while safety is a prerequisite for participation, it does not guarantee integration. Broader CSCW scholarship establishes that successful community entry requires newcomers to acquire specific technical and cultural competencies~\cite{kraut2012building, wenger1999communities}. In the specific context of Social VR, the user-driven strategies required to navigate the combined technical, usability and normative friction of onboarding remains under-theorized. This study addresses this gap by shifting the analytical lens from platform features to the lived adaptation processes of the newcomers themselves.}

\subsection{\abnew{Newcomers’ Roles and Challenges in Online Spaces}}
Attracting and retaining newcomers is essential for the sustainability of online platforms. Newcomers compensate for member turnover and introduce fresh perspectives necessary for community evolution~\cite{Kiene356, Mitra2013Effect}. \abrev{However, this initial phase is characterized by extreme fragility. Empirical studies across diverse platforms, from wikis~\cite{panciera2009wikipedians} to open source projects~\cite{ducheneaut2005socialization} and online communities of interest~\cite{Karumur2016Early}, consistently show that the majority of users disengage after their very first interaction. Consequently, research has increasingly focused on the friction points that impede integration, shifting the question from why newcomers join to how they survive this precarious transition.}

\abrev{To conceptualize this transition, we draw on the framework of \textit{Legitimate Peripheral Participation} (LPP)~\cite{lave1991situated}, which reframes onboarding not as a transfer of information~\cite{kraut2012building}, but as a process of negotiated entry. LPP posits that newcomers learn by observing and mimicking ``old-timers'' from a safe periphery before gaining the competence to participate fully. However, in complex online environments, this path is rarely linear. As Wenger notes, large communities often function as a \textit{Constellation of Practices}~\cite{wenger1999communities}, where users must navigate a fragmented landscape of sub-cultures rather than a single monolithic group. For newcomers, the challenge is securing legitimate access to these interlocking communities despite lacking local cultural capital.}

\abrev{While much research has examined these dynamics in task-oriented platforms (e.g., Wikipedia~\cite{antin2010readers}, GitHub~\cite{dabbish2012social}), challenges in open-ended social spaces are distinct. In these environments, barriers are not just procedural but deeply social and spatial.}

\abrev{In open-ended virtual worlds, newcomers often face immediate social stratification. Earlier work in text-based virtual environments, such as LambdaMOO, documented steep learning curves and ambiguous norms \cite{dibbell1994rape, reid1994cultural}. These communities developed onboarding rituals and roles to scaffold participation~\cite{isbell2006cobot}. Later research on graphical virtual worlds, such as Second Life (SL), revealed similar struggles~\cite{boellstorff2015coming, malaby2011making}. Ethnographic work on SL highlights the ``stigma of the newbie,'' where veteran users actively identify and marginalize newcomers based on visual markers like default avatars or clumsy behavior~\cite{boostrom2008social}. Unlike text-based forums where a user can lurk invisibly~\cite{nonnecke2000lurker}, the visual nature of virtual worlds makes the newcomer's lack of cultural capital immediately visible, leading to potential exclusion before they can successfully mimic established norms~\cite{boostrom2008social, malaby2011making}. Just as supportive replies are crucial in text communities~\cite{lampe2005follow}, the lack of social scaffolding in virtual worlds can make the environment feel hostile and impenetrable.}

\abrev{In 3D environments, social competence relies on technical proficiency. Newcomers must simultaneously master movement, camera control, and personal space~\cite{locher2015negotiation}. Technical errors, such as walking into walls, signal incompetence and disrupt the community's immersion~\cite{locher2015negotiation}. Unlike MMOGs where game mechanics guide interaction, the lack of predefined goals in social worlds leaves newcomers without clear direction on how to behave, increasing the cognitive load of adaptation~\cite{messinger2009virtual}.}

\abrev{Although historical virtual worlds faced similar challenges, Social VR complicates this peripheral learning process through embodied presence~\cite{slater1997framework}. The avatar acts as a persistent proxy for the user's attention and presence~\cite{yee2007unbearable}. Consequently, even passive observation is socially legible to others; as Locher et al. note, simply standing in a virtual space is an interactional act that requires managing proximity and gaze~\cite{locher2015negotiation}. The specific strategies users employ to navigate this high-exposure adaptation remain underexplored, a gap this study aims to address.}

\section{Research Method}
We conducted a qualitative study to investigate the difficulties newcomers face in social VR and the strategies they employ to adapt. To do so, we interviewed VRChat users and analyzed the data using thematic analysis.

\subsection{\qj{Selecting VRChat as a Case Study}}
Social VR is accessible through several platforms. We choose to focus on a single platform after thorough considerations.
\qjnew{The choice is grounded in the literature. Case studies are an effective method for examining complex social phenomena in depth~\cite{yin2018case}.} A focused approach allows for a deeper analysis of user behaviors and dynamics. This depth supports theory-building by revealing underlying mechanisms and generating insights that can apply to other similar contexts \cite{eisenhardt2007theory}. \qjnew{Practically}, resource-intensive qualitative methods, such as interviews, ethnography, and content analysis, require concentrated effort and detailed engagement, which is less feasible across multiple platforms without sacrificing analytical depth \cite{stake1995case}.
Single-platform studies are also a well-established approach within the HCI and CSCW research communities. Platforms like Twitch have been extensively studied as representative environments for live streaming \cite{sheng2020virtual, wu2023interactions}, and League of Legends has often served as a model for MMORPG gaming research \cite{kou2021flag, kou2018playing}. In broader studies of newcomers, single-platform case studies are also common; for example, Facebook has been used as a case study for generating insights into social networking sites \cite{burke2009feed}. 

VRChat was chosen specifically for the following reasons. First, VRChat is one of the most popular applications on the Steam store, attracting a diverse and heterogeneous user base \cite{maloney2021social}. This diversity provides a rich context for analyzing the challenges and strategies of newcomer integration. Furthermore, VRChat has garnered significant academic interest due to its scale, openness, and popularity, making it an established research subject \cite{deighan2023social, Kexue, Chen05}. This validates its relevance as a context for in-depth exploration of unique challenges and behaviors in social VR.

Although social VR platforms potentially vary in terms of community culture, moderation policies, and technical implementations, they are built upon a common set of core affordances, such as embodied avatars, spatialized audio, and open-ended environments. These shared features shape how users interact, form relationships, and navigate the platform. By focusing on VRChat, we engage with these affordances in a rich and widely used environment, offering insights relevant to the design and user experience considerations present across immersive social platforms

\subsection{\qj{Participants}}
Participants were recruited through different methods, including disseminating recruitment messages via the researchers' networks (n=6), sharing calls for participation across online forums, such as Discord (n=10), and snowball sampling (n=8). 
\qj{In defining newcomers, researchers have relied on objective metrics such as time since joining or registration, or total activity (e.g., number of posts) \cite{gallagher2015becomes, rollag2004impact, burke2009feed}. However, setting thresholds (e.g., defining newcomers as users within their first year and old-timers as those beyond it) can oversimplify the complex, evolving nature of newcomer experiences and community engagement. The process of integration or socialization is likely to be continuous, with no clear threshold separating newcomers from old-timers \cite{comer1991organizational}.} 
\qj{To address these nuances, we allowed participants to self-identify their newcomer status based on their own experiences \cite{Singh2012Newcomer}. In our recruitment, we specifically invited individuals who identified either as newcomers or as experienced users of social VR, framing our study as an exploration of newcomer integration challenges. 
Involving users at different stages of engagement allows for richer insights. Experienced users, for example, can provide valuable perspectives based on their own journeys. Their deeper understanding, developed through observing and interacting with newcomers, enables them to articulate barriers and challenges with greater clarity.
In fact, some studies deliberately involve only experienced users when investigating newcomer experiences. For instance, \citet{balali2018newcomers} recruited 10 experienced users for interviews to explore challenges in newcomer interactions and barriers to integration.}
Ultimately, we recruited 13 newcomers and 11 experienced users for this study. The sample consisted predominantly of male participants (n=14), along with 8 females, one non-binary individual, and one participant who chose not to disclose their gender. All participants reported prior experience with VRChat, with additional experience spanning platforms such as AltspaceVR and RecRoom. Further demographic and engagement details are available in \autoref{tab:interviewee}. Here, we use VRChat as a specific environment to elicit participants' inputs for understanding the newcomers' challenges and integrations.
We did not include individuals who tried the platform only once or twice before leaving. 
Although this group could provide insights into immediate deterrents, our study focused on newcomers who had actively attempted to engage with social VR and could therefore reflect on both the difficulties they encountered and the ways in which they navigated early integration. In addition, experienced users often observe newcomers who quickly disengage, and thus can indirectly articulate the types of difficulties that may drive early dropout. More, conducting in-depth interviews requires participants to have a sufficient experiential basis to reflect on their trajectories and the challenges they faced. By recruiting self-identified newcomers and experienced users, we ensured that participants could provide detailed accounts of both early struggles and longer-term adaptation. This approach enabled us to maintain analytical depth while capturing a broad spectrum of integration experiences.

\subsection{\qj{Interviews}}

The interviews ranged from 32 to 71 minutes, with a median of 44 minutes and an average of 42 minutes. \final{This variation is typical in semi-structured interviews, where the depth of responses, the need for follow-up questions, and the complexity of participants’ experiences naturally differ across interviews. All interviews followed the same protocol, but some participants provided more detailed narratives or raised issues that warranted additional probing.} We conducted the interviews either via voice chat software outside VRChat or through avatars within VRChat, depending on participants’ preferences. In the latter case, interviews were conducted in a private social VR space with only the interviewer and interviewee present, in order to avoid disruption and protect participants’ privacy. Each participant received a €20 Steam gift card in recognition of their time and contribution.

\abrev{After establishing the participants' backgrounds and general usage context, the interview protocol focused on their integration journey.} 
We started by asking participants to describe their first experience with social VR. For example, we asked: "Can you describe your first experience with social VR?" This helped us gather initial information about their overall experiences and set the stage for a more detailed exploration.
Depending on their responses, we either delved deeper into their answers (e.g., if their answers related to negative experiences or directly mentioned the difficulties newcomers face) or transitioned into more specific inquiries about the challenges they faced as newcomers. For example, we asked, "What challenges do you experience as a newcomer in general?"
Based on their responses, we probed further into the difficulties they identified. If participants struggled to pinpoint specific challenges, we prompted them with more detailed questions related to technical issues and social aspects. For example, `What technical issues did you face when first using social VR?' What difficulties did you experience in navigating the virtual environment? `Did you face any challenges in communicating with other users? If so, what were they?' 
After discussing the challenges, we shifted our focus to exploring the mechanisms through which newcomers acclimate and integrate into social VR. We asked participants questions designed to uncover their adaptation processes, such as: `How did you familiarize yourself with the VR environment?' `What steps did you take to integrate into the community?'
Based on their responses, we posed additional questions to gain more detailed insights into their acclimation and integration strategies. For example: `What resources did you use to support your integration?' `Did you receive any help from anywhere? If so, can you elaborate?'
This semi-structured approach allowed us to comprehensively explore the initial challenges faced by newcomers and the strategies they employed to overcome these challenges and integrate into the social VR environment.

\subsection{Data Analysis}

We utilized reflexive thematic analysis \cite{braun2021one, braun2022conceptual, braun2019reflecting} to understand the complexities of newcomer integration in virtual environments. Reflexive thematic analysis excels at uncovering both explicit and implicit nuances in qualitative data, making it ideal for our study.

\begin{figure}[h]
    \centering
    \includegraphics[width=1\linewidth]{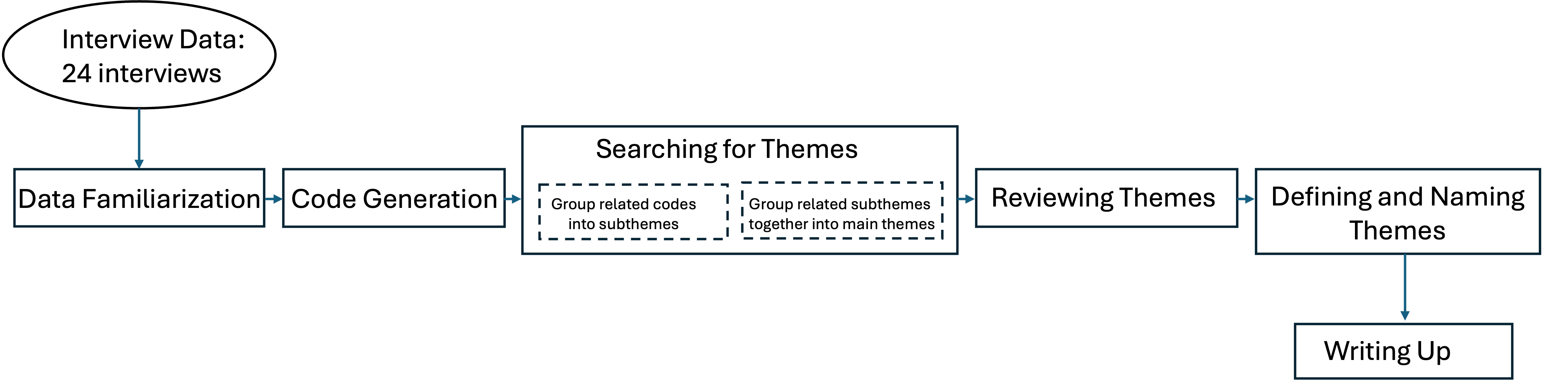}
    \caption{Flow of the Data Analysis}
    \label{fig:2}
\end{figure}

Two researchers were involved in developing the themes through steps outlined by Braun and Clarke \cite{braun2006using, braun2019reflecting}, \qj{as shown in \autoref{fig:2}}. The analysis began with an in-depth immersion in the dataset to develop a comprehensive understanding of the data and its context. Initially, both researchers independently read through the collected data, noting their initial impressions and identifying emerging patterns. They then met to discuss their preliminary understanding and confusion in certain contexts of the data before returning to the dataset to generate initial codes independently.
We employed an iterative coding process, where codes were generated in a data-driven manner. Each piece of data was scrutinized, and segments of text were coded for their underlying meanings. For example, a comment such as `Newcomers can easily get motion sickness as they not know certain actions can cause it... Many of them like to do things like running fast or frequently turning' was initially coded as ‘Unaware of causes of VR sickness’.
Through regular meetings, we discussed and refined our codes, grouping them into potential themes. For example, codes like ‘Unaware of which behaviors causing VR sickness’ and ‘Unaware of which environments causing VR sickness’ were grouped under the broader theme of ‘Lack of knowledge on avoiding VR sickness’. This iterative process involved continuous back-and-forth between the dataset and the emerging themes to ensure they accurately reflected the data.
Next, we examined internal relationships within sub-themes to develop overarching themes. This step involved identifying patterns and connections across various sub-themes to create a cohesive understanding. We critically assessed these themes in light of our research objectives, evaluating their relevance to understanding newcomer integration in the VRChat environment. This required multiple iterations of reviewing the data and themes to ensure a comprehensive and precise depiction of users’ experiences. In line with the principles of reflexive thematic analysis \cite{braun2021one} and the views expressed by McDonald et al. \cite{mcdonald2019reliability} regarding the role of codes in research as part of the analytical process rather than the final outcome, we deliberately did not pursue inter-rater reliability. 

\subsection{\qj{Positionality}}

\qj{Positionality statements encourage researchers to critically reflect on their own biases, identities, and assumptions, which are inevitably shaped by their lived experiences \cite{bourke2014positionality, corlett2018reflexivity}. This reflexivity ensures a mindful approach to the research process, fostering greater transparency and trustworthiness in the study's findings. All the authors of this study bring prior experience with social VR through personal interest and engagement in previous research endeavors. Specifically, the authors have had meaningful contact with newcomers to social VR, for example, engaging with them face-to-face in both virtual and physical contexts as part of prior research efforts. These experiences provide valuable insights into the challenges newcomers face when navigating social VR spaces.
The first author, in particular, has several years of experience across a variety of social VR platforms, having the full journey from being a newcomer to becoming a seasoned participant.}

\section{Findings}

\subsection{\abrev{Barriers to Integration}}
This section highlights the key challenges newcomers encounter in social VR platforms, including adapting to unfamiliar social norms, navigating open-ended environments, coping with VR sickness, learning new functions and user interfaces, and overcoming socialization difficulties. Notably, VR sickness poses a greater challenge for newcomers who have no prior experience with VR technology before using social VR.

\subsubsection{\abrev{Navigating Implicit and Embodied Social Norm}}
The social norms that form on social VR platforms pose unique challenges for new users. These norms are uncommon in other online or offline settings, imposing difficulties to newcomers and requiring newcomers to spend time adapting to these differences. \qj{For instance, 
in the environments, gestures like virtual physical touches, such as pats, head touches, or hugs, are commonly used in social interactions even among strangers \cite{maloney2020talking}. These gestures, though entirely virtual, are used by many to simulate physical intimacy and emotional connection.} However, for new users unfamiliar with such interaction in virtual spaces, it might be uncomfortable or feel unnatural to new users who are accustomed to traditional social boundaries. For example, \textit{
    ``Many players use a lot of touches to communicate with others. New players tend to really struggle with accepting this at first. They can feel somewhat 
    intimidated by such interactions. I've seen that quite often.'' (P18)}
\qj{P3 echoed the response, stating, \textit{``Many people here like to make physical contact, like patting each others' heads, touching some parts of bodies, and stuff like that. A lot of people do it... but I’m still not really used to it. It feels kind of weird.''}}

Another example is role-playing. Role-playing and the exploration of alternative identities are prevalent. They are significant motivations for many users in social VR \cite{freeman2021body,15Acting}, \qj{For instance, many transgender players use avatars of alternative genders to better facilitate their identity expression \cite{15Acting}. The disguise in social VR (encompassing both voice and appearance}) is not typically encountered in other face-to-face interaction settings such as in the physical world.
\qj{For seasoned users, this flexibility is a source of creativity and entertainment. However, newcomers are often unaware of the prevalence and depth of role-playing. For example, }

\begin{quote}     \textit{
    ``Many new players gravitate around users who use female avatars with appealing voices, but that may be male players... However, the newcomer might not be aware of this... They don't know that this kind of role-playing is very popular. They tend to take it for granted and then really get into it.'' (P5)
} \end{quote}

\qj{New players, unfamiliar with the cultural norms, may not realize the extent to which role-playing shapes online interactions. They tend to take the virtual personas at face value, assuming a direct correspondence between the avatar's identity and the real-world individual controlling it. }

The misunderstandings can lead to feelings of betrayal and mistrust when the reality of role-playing becomes apparent. It can make them feel deceived and doubtful about the authenticity of the social experience. 
 For example, \textit{
    ``Those experiences negatively impact their experience. Many might even start feeling that everyone there is fake, and all the interactions are fake.'' (P18)}

\subsubsection{\abrev{Disorientation from Unstructured Open Worlds}} 
\qj{Social VR platforms are designed as open-ended environments, prioritizing user-driven interaction over structured gameplay or predefined goals \cite{wiki:VRChat, kolesnichenko2019understanding, maloney2021social}.} 
\qj{Social VR platforms often leave users to define their own purpose \cite{sykownik2021most}.} 
\qj{The absence of clear, structured gameplay presents significant challenges for newcomers, particularly around user retention and meaningful engagement. }
For example, 

\begin{quote}     \textit{``The biggest problem is probably not knowing where to go or how to play the game... it's not like some normal games where you fight monsters to level up. So the retention rate for new players is actually very low. If 2 out of 10 people stay, that's already considered high...''} \end{quote}

\qj{For new users accustomed to the structured progression of typical games, this open-ended design can be disorienting. According to P6, this lack of structure leads to uncertainty, with newcomers unsure of how to spend their time or derive value from the platform. The open nature of the environment becomes a barrier rather than an opportunity, particularly for users who prefer more guided experiences. } 

Experienced users primarily engage in social activities with friends. However, newcomers struggle as they do not have or have not yet established a social network within these platforms. This lack of social connections, in combination with no direction, leads to frustration and potential disengagement from the platform. For example, 

\begin{quote}     \textit{``It doesn’t have a specific theme like other games, where what you're supposed to do is very clear. It's a very free and open community. No matter what kind of world or how you play, ultimately, it relies on a group of people for social interaction. A major reason why many newcomers don't stay long is that they do not establish their social networks on the space. (P11)
''} \end{quote}

\qj{Without clear objectives or immediate social ties, newcomers often feel isolated. This isolation compounds their difficulty in navigating the platform and creates a sense of purposelessness. As a result, many users leave before fully exploring the platform's potential.}

\subsubsection{\abrev{Physiological Barriers to Social Presence}}
\qj{VR sickness, a form of motion sickness triggered by immersive virtual environments, includes dizziness, nausea, and discomfort \cite{chang2020virtual, chattha2020motion}. It presents a significant challenge for newcomers to VRChat, particularly those without prior experience with VR technology. The symptoms can severely impact new users' initial experiences, reducing their willingness to engage further with the platform. For example, }
\textit{``VR sickness is a big problem. It can be quite challenging for newcomers.’’ (P4)}
\qj{New users are particularly vulnerable to VR sickness due to their lack of exposure to the sensory and physiological adjustments required for VR. Unlike experienced users who have acclimated to the medium, newcomers often experience more intense symptoms. As P22 described,}

\begin{quote} 
    \textit{``For beginners, playing VR games can easily lead to motion sickness, especially since they have not previously experienced these symptoms. Many veteran users I know don't get dizzy. They just get used to it and can often play for several hours at a time.''} \end{quote}
    
\qj{Over time, frequent exposure can reduce susceptibility to VR sickness, but this requires persistence, which many new users may lack due to their initial negative experiences.}

In addition, the lack of knowledge among new users about how various behaviors or settings, such as the choice of worlds, can further exacerbate their experience. \qj{The different settings can impose different levels of sickness on users, such as the speed of movements, evenness of terrain, or visuals \cite{chang2020virtual}}. New users often lack awareness and thus do not have the consciousness to avoid the conditions. For example,  

\begin{quote}     \textit{``It's related to the environment. Some environments can be particularly disorienting. Newcomers often don't know which environment is less likely to cause dizziness, which definitely can make their experience more uncomfortable.'' (P7)}\end{quote}

\qj{The combination of physiological sensitivity and limited knowledge about mitigating factors leaves newcomers more prone to discomfort. Experienced users can adjust settings, avoid triggering environments, or adopt techniques (e.g., focusing on fixed points or limiting session times) to manage VR sickness. Such strategies are often unknown to novices.}

\subsubsection{\abrev{Cognitive Overload from Spatial Interfaces}}
The novel functionalities and UI in social VR platforms present unique challenges and learning curves for new users. Social VR offers a different experience from traditional online platforms due to its immersive and embodied nature. \qj{Many features are not merely adaptations of existing 2D platform functions but are innovative designs tailored for the 3D immersive environment \cite{mcveigh2019shaping}.}
For example, audio controls that allow users to manage sound settings extensively—such as muting specific areas or focusing on conversations within a certain group—are tailored to enhance social interaction within noisy environments. These functions are beneficial and often crucial, but many users are unaware of their existence, do not know about their effects, or how to use them.

\begin{quote}     \textit{``For example, you might not have noticed, there is a feature like muting sound within a specific area. If I find the people around me too noisy and I just want to listen to the conversation within my immediate circle, I can turn on this feature... Many features are quite innovative, but newcomers do not know.''} (P21) \end{quote}

\qj{The novel features can be observed across different platforms. For example, Rec Room take advantage of VR’s 3D environment and hardware to create intuitive interactions. For instance, players can shake hands with other avatars by moving their controllers and slightly pulling the trigger—an embodied action that facilitates social bonding. In England, the introduction of a ``panic button'' provides users with an immediate and innovative way to escape from harassment to tackle social challenges \cite{mcveigh2019shaping}.}

Unlike in traditional online environments, where knowledge can be somewhat transferable between similar games or other similar online virtual spaces, social VR requires acclimation to a different set of UI. For example, 

\begin{quote}     \textit{``You need to learn from zero. The UI is quite unique... Most other games are played on a PC with a flat-screen, but social VR like VRChat is different. It differs from other online platforms I'm familiar with... It's specifically designed for VR devices, quite distinctive.
''} (P11)\end{quote}

\qj{The comparison to other games or platforms further emphasizes this uniqueness. Social VR is described as ``different'' and ``quite distinctive,'' suggesting that even seasoned gamers or users of online platforms must undergo a learning period. }

\subsubsection{\abrev{Vulnerability of Embodied Visibility}}
As mentioned in the previous subsection social interaction is central to the experience of these platforms. 
However, many new users struggle to establish their own social networks within the platform. 
For example, \textit{``If you want to make friends there, the first problem you encounter as a new user is that you don’t know how to make friends.'' (P22)}
Even the act of speaking or greeting, which may seem trivial in conventional online social settings, poses a hurdle in social VR,
    \textit{``It's really common for new players not to use voice chat... When I first started using VRChat, I went at least a hundred hours without speaking to anyone.'' (P18)}

The heightened social anxiety contributed to users' inhibition to initiate or sustain social interactions. 
\qj{The anxiety in social VR is potentially exacerbated by the immersive nature of VR settings \cite{zamanifard2023social}.
Spatial audio, expressive avatars, and the sensation of social and physical presence make VR interactions feel more immediate and intense. While these elements make VR interactions engaging, they can also amplify feelings of social pressure. 
This anxiety is particularly challenging for newcomers, who often feel significant pressure to build social connections. Many participants in this study emphasized that speaking up in social VR is more intimidating than in other online spaces.} For instance,

\begin{quote}     \textit{
    ``Talking to people in social VR can be more challenging. When gaming and using voice chat, it's like talking on the phone, so you can speak freely without much pressure. But in VRC, since everyone is using VR, it's more like real-life socializing with face-to-face communication... It can definitely feel more difficult...'' (P15)}\end{quote}

\qj{In addition, interacting with unfamiliar individuals can intensify feelings of discomfort and hesitation \cite{duronto2005uncertainty}. Newcomers often need to interact with strangers to establish friendships, which exacerbates their social anxiety.} P1 likened this experience to a real-world scenario,
\textit{
     ``It is kind of like striking up a conversation with a stranger on the street... Most people may not be social enough to strike up conversations or try to make new friends on the street.''}
\qj{This comparison underscores the heightened difficulty of forming new relationships in VR, where the immersive environment mirrors the social pressures of real life.}

 Given the conditions, if new users seek interaction and guidance but receive a lukewarm response or no response at all from veteran users.
 It can further discourage them \qj{from socializing}. For example,
 \textit{ ``For new players, not receiving any response when they first use voice chat can be quite disheartening. It might even make them reluctant to use it again for a while.'' (P14)}
\qj{This lack of acknowledgment can leave new users feeling ignored and unwelcome, significantly diminishing their willingness to engage further.}

\subsubsection{\abrev{Navigating Social Gatekeeping and Insularity}}

Besides struggling with effective socialization, many new users in social VR find it difficult to integrate due to the exclusivity often displayed by more experienced users.

\qj{New users often lack knowledge about game mechanics and social rules of the immersive virtual world.
As we discussed before, their unfamiliarity with existing practices can result in poor experiences for themselves. The inexperience can also disrupt the interaction flows of others, inadvertently frustrating veteran users. As a result, established users may avoid engaging with newcomers, preferring to minimize potential misunderstandings or disruptions.}

\qj{The contrasting interests of new and experienced users further reinforce the divide. Newcomers often exhibit enthusiasm for exploring maps, avatars, or experiences that veterans have already exhausted. This divergence can make collaboration or shared activities less appealing for veteran users. For example,}

\begin{quote}     \textit{``
    For veterans, we have already experienced everything. I don't have the desire to revisit those beginner experiences. When I first started, I found things like exploring new maps and playing horror games fascinating. Now, it just doesn't interest me the same way; I'd rather just hang out with my old friends.
''} \end{quote}

\qj{As expressed by P19, this disconnect can lead veterans to prioritize interactions with peers who share their level of familiarity and interests. This unintentionally leaves newcomers excluded.}

Veteran users often resist integrating new members into their established social circles due to the perceived effort and disruption involved. Adding a new user requires adjustments in group dynamics, the re-explanation of norms, and potentially a reconfiguration of the group's activities. This additional effort is seen as a social cost. As explained by P10,

\begin{quote}     \textit{``Bringing in someone from outside can mess with the dynamic. It feels like the space is being invaded. It's like there's this huge social cost to it. A lot of folks, once they've already got a bunch of friends, they tend to stick to private maps, keeping things pretty exclusive...''} \end{quote}

\qj{This tendency for groups to maintain exclusivity and stick to private spaces further alienates newcomers, limiting their opportunities for inclusion.}

Given the above, even when veteran users are open to expanding their circles or need to interact with others, they often prefer to connect with users who are similar to themselves in terms of experience or interests. For example, \textit{``Even if experienced users are open for friends, they're more likely to find others who have similar experience. They might not be keen to include new players.'' (P3)}

In the previous section, we mentioned that if veteran users do not respond well, it can be discouraging for new users. This behavior by veteran users creates a barrier to entry, compounding the difficulties new users already face in navigating and understanding the dynamics of the VR platform. For example, \textit{``The exclusion issue I mentioned is actually pretty crucial, and it kinda creates a vicious cycle, you know?..It totally ruins the experience.''(P24) } 
\qj{Similar to findings in knowledge-sharing platforms, an unrewarding first attempt can discourage further contributions or interactions \cite{steinmacher2012recommending,Morgan2018Evaluating}. In social VR, this phenomenon can be particularly damaging as it directly undermines the central appeal of the platform—forming connections and engaging with others.}

\subsection{How They Familiarize Themselves to the New Environment}

The section outlines how users familiarize themselves with the novel online space, focusing on two aspects: technical operations and social norms. It is structured into two sections, 4.2.1 covers technical aspects, such as how to use specific features of the platform. Section 4.2.2 addresses social norms, discussing how users can understand and adapt to the behavioral expectations within the platform. 

\subsubsection{\abrev{Self-Directed Technical Adaptation}}
Users of social VR platforms typically acquire the technical skills required to navigate and utilize. These learning processes involve exploring the platform independently, leveraging online resources, and seeking guidance from veterans. However, the effectiveness and accessibility of these methods vary, often leaving gaps in the user experience.

\qj{\textbf{Leveraging External Resources.}}
Additionally, many users utilize online resources to gather information and learn. These can include forums and video tutorial sites. This helps new users to understand and master the platform more efficiently. 
For example, 
\textit{
     ``There are videos and tutorials online. Like, there are lots of videos on YouTube, that can provide support for newcomers to get started... so you can enhance your understanding of the platform through those as well.'' (P22)}
\qj{This comment underscores the importance of accessible, user-friendly external resources in helping newcomers bridge the knowledge gap. The availability of tutorials enables users to overcome specific challenges more efficiently, reducing the learning curve.}

\qj{\textbf{Through Trial and Error.}}
\qj{Self-learning is one of the most common ways users adapt to the technical aspects of social VR. Through repeated exploration and trying of features, users gradually familiarize themselves with the platform's tools and functionalities, often over an extended period. For example, one participant described this process,}
\textit{``It's because of long-term use, like searching the web or finding friends through search functions. Since you use them often, eventually you get the hang of it.'' (P7)} 
Despite self-exploration being a common practice, there is still a significant demand for more structured guidance. 
Users desire instruction or mentors who can provide a personalized introduction to the platform, helping alleviate potential frustration and reduce the learning curve. For example, 
\begin{quote}
\textit{``You still need reliable instructions to show you the space, or someone willing to guide you into it. Sure, you can figure it out on your own, but it might take forever, and the whole process can be pretty annoying and frustrating.'' (P4)}
\end{quote}

\qj{While learning self-exploration fosters independence, it can be overwhelming without clear direction. The absence of formal onboarding mechanisms often leaves newcomers feeling unsupported as they navigate the platform's complexities.}

\qj{\textbf{Inquiring Experienced Users.}}
Beyond self-exploration, actively communicating with experienced users and seeking their advice is another crucial method. \qj{This approach is particularly beneficial for understanding specific operations or navigating complex features that may not be intuitive through trial and error. Engaging with veteran users provides newcomers with practical tips and insights, significantly easing their adaptation process.} This type of interaction is especially common when newcomers have pre-existing relationships with experienced users. For example, \textit{``Some of my friends, who are also relatively new users like me, get guidance from a few experienced users. As we often hang out with them. So we often turn to those users for help.'' (P17)}

However, while guidance from experienced users is highly beneficial, it is not always readily available or sustainable over the long term. Many times, users must rely primarily on their own efforts to solve problems. \qj{Assistance from veterans tends to be sporadic and situational.}  For example,

\begin{quote}
\textit{``Another way is through guidance from experienced users, but most of the time, users still have to figure things out on their own. Some experienced players might be willing to help newbies, but it's usually just once in a while. For example, if someone doesn't know how to use the camera, a veteran player might show them which button to press. But for long-term help, I haven't seen much of that happening.'' (P21)}
\end{quote}

\qj{This quote underscores the limitations of relying on experienced users as a consistent source of support. While their assistance can be invaluable for addressing specific issues, it rarely extends to providing comprehensive or ongoing guidance.}

Additionally, many users turn to external platforms to learn from experienced users. This can involve participating in online forums or engaging in discussions on platforms like Discord.

\subsubsection{\abrev{Social Calibration}}
When learning social norms on the platform, the process also involves the above two ways, but it partially differs from learning technical attributes.

\qj{\textbf{Experiential Learning.}}
\qj{Many users initially find some norms or social etiquette in the social VR environment confusing. They do not understand why certain actions are performed or encouraged. However, as they actively engage with the platform and try these actions firsthand, they begin to comprehend the reasons behind these social norms. This direct engagement helps them adapt to the community's expectations and feel more comfortable with its dynamics.}
For example, P10 states, 
\begin{quote} \textit{``When I first started as a new player, I didn't quite understand why most people like to hang out in front of mirrors or do other things. But after trying it out myself, I started to understand it a bit more. I gradually got used to it, and it wasn't something particularly difficult to accept. It's just that I learned and understood it through doing.''} \end{quote}

\qj{This example highlights the phenomenon of ``Mirror dwelling'', or the practice of spending time in front of virtual mirrors. While it may initially seem peculiar to newcomers, experienced users regard it as a relaxing and social activity that fosters connection and self-expression \cite{Kexue}.}

\qj{Experiential learning in social VR is not limited to users’ own actions. Being affected by the behavior of others also plays a crucial role. For example, as stated by P5,} \textit{``Many new players might not get why some of their actions can annoy others. But as they spend more time playing and feel the irritation of being affected by others doing the same things, it starts to make sense...''}
\qj{Encountering inappropriate or disruptive actions by others can help users understand the importance of adhering to community norms. Through these interactions, they come to appreciate not just what to do, but also what to avoid, thereby contributing to a more harmonious social environment.}

\textbf{\abrev{Peripheral Observation and Strategic Mimicry.}}
A common way newcomers adapt to social norms in virtual environments is by quietly observing and imitating experienced users. This approach enables them to internalize the subtleties of social etiquette without the pressure of direct interaction, which can be particularly helpful when they are uncertain about how to behave. By closely watching how seasoned users navigate various social scenarios, newcomers can adopt these behaviors more naturally and effectively.
P14 emphasized the significance of observation,
\begin{quote} \textit{``Most of them just stay quiet, watching and hanging around without saying many words. They observe how experienced users play and communicate... They’re just learning from the experienced, mimicking their actions, and following.''} \end{quote}

\qj{Mimicking experienced users provides a safe and low-risk way for newcomers to experiment with their social engagement. By adopting these observed behaviors, they gain confidence and gradually find their place within the community. This method of passive learning often serves as a stepping stone to more active participation, where newcomers transition from observers to full-fledged contributors within the social environment.}

\qj{\textbf{Feedback Through Interaction.}}
\qj{Feedback from others plays a crucial role in helping users refine their understanding of social norms. This feedback can be direct, such as receiving comments or corrections.  It help users identify what is acceptable and what is not within the community.
For instance, P6 user explained, 
\begin{quote} \textit{``When I first started using VRChat, I didn’t know that using visual effects was not allowed. After doing it a few times, someone told me that it was wrong, and that’s when I realized... In VR, many players sleep, and visual effects can negatively impact some players, such as those with epilepsy.''} \end{quote}}
This illustrates how direct feedback can educate users about the consequences of their actions and the reasons behind certain restrictions.

\qj{In addition, this feedback can be indirect, it involves observing how others react to a particular behavior or situation and using these reactions to infer what is acceptable. By paying attention to the responses of others, users gain insights into the community's unspoken norms.
For example, P22 explained,} \textit{\qj{``At first, I didn’t know what was considered annoying. But by seeing how others reacted to certain actions—whether they laughed, got irritated, or stayed quiet—I just started to understand what to avoid.''}}
\qj{ This indirect form of feedback helps users pick up on subtle rules that are not explicitly stated but are essential for harmonious social interactions in VR.}

\qj{\textbf{Social Assimilation.}} \qj{While experiential and observational learning are key steps, the broader process of social adaptation occurs over time as users internalize the community's norms and practices. Through repeated exposure and participation, users gradually align their behavior with the community's expectations. With time, they unconsciously adopt these norms.} \qj{For example, \textit{``I felt like as I spent more time on the platform, I started to act similarly without even thinking about it—it just became natural.'' (P8)}} \qj{ As users participate in activities, engage with others, and adapt to the platform’s unique social dynamics, they move from feeling like observers to becoming integrated members.}

\qj{Similar to learning technical aspects, some users independently research to better understand social norms. They may seek out articles, discussion threads, or guides that explain the etiquette and unspoken rules of the platform. This self-directed learning complements the other methods and helps users gain an understanding of social expectations. Additionally, some users actively learn through direct inquiry from the experienced.}

\begin{table*}[ht]
  \centering
  \caption{\abrev{The newcomer persistence temporal framework. This table maps the specific barriers identified in our findings, the adaptation strategies used by survivors to overcome them, and the corresponding design interventions proposed to support integration.}}
  \label{tab:persistence_framework}
  \small
  \renewcommand{\arraystretch}{1.5} 
  \begin{tabularx}{\textwidth}{@{}l>{\raggedright\arraybackslash}X>{\raggedright\arraybackslash}X>{\raggedright\arraybackslash}X@{}}
    \toprule
    \abrev{\textbf{Stage}} & \abrev{\textbf{Barriers}} & \abrev{\textbf{Adaptation Strategies}} & \abrev{\textbf{Design Implications}} \\
    \midrule

     \abrev{\textbf{1. Acclimatization}} & 
    \abrev{\textbf{Physiological and Technical Friction}} \newline
    \abrev{Newcomers face barriers including VR sickness due to sensory conflict (Sec. 4.1.3) and cognitive overload from complex spatial interfaces (Sec. 4.1.4).} & 
    \abrev{\textbf{Trial and Error and External Resources}} \newline
    \abrev{Users rely on self-directed trial and error to master tools (Sec. 4.2.1). When stuck, they leverage external resources (e.g., YouTube tutorials) or seek sporadic help from experienced users (Sec. 4.2.1).} & 
    \abrev{\textbf{Systemic Safety and Scaffolding Interactions}} \newline
    \abrev{Implement comfort-first defaults (e.g., teleportation, reduced FOV) and "ghost visualizations" to teach spatial controls via bare-hand guidance (Sec. 5.4.1).} \\ \addlinespace

    \abrev{\textbf{2. Acculturation}} & 
    \abrev{\textbf{Normative Ambiguity and Exposure}} \newline
    \abrev{The hidden curriculum of norms creates confusion (Sec. 4.1.1). The embodied nature of VR prevents invisible lurking, causing performance anxiety (Sec. 4.1.5).} & 
    \abrev{\textbf{Observation, Mimicry and Feedback}} \newline
    \abrev{Survivors adapt by quietly observing and mimicking veterans to internalize etiquette (Sec. 4.2.2). They refine behavior through feedback and interaction, utilizing direct corrections and indirect reactions (Sec. 4.2.2).} & 
    \abrev{\textbf{Spatial Buffers and Normative Visualization}} \newline
    \abrev{Create designated observation decks to legitimize lurking (LPP) and use simulation-based tutorials to distinguish between playfulness and harassment (Sec. 5.4.2).} \\ \addlinespace

     \abrev{\textbf{3. Embedding}} & 
    \abrev{\textbf{Purposelessness and Isolation}} \newline
    \abrev{The lack of predefined narratives or quests leads to disorientation (Sec. 4.1.2). Users struggle to move from superficial interactions to deep social bonds (Sec. 4.1.5).} & 
    \abrev{\textbf{Social Assimilation and Experiential Learning}} \newline
    \abrev{Users achieve persistence through social assimilation, unconsciously adopting norms over time (Sec. 4.2.2). They engage in experiential learning (e.g., mirror dwelling) to understand the social value of community rituals (Sec. 4.2.2).} & 
    \abrev{\textbf{Intelligent Scaffolding and Social Bridging}} \newline
    \abrev{Deploy AI-based embodied agents for context-aware orientation and gamified objectives (e.g., "Community Passports") to structure exploration and foster mentorship (Sec. 5.4.3).} \\
    
    \bottomrule
  \end{tabularx}
\end{table*}

\section{Discussion}
In this section, \abrev{we synthesize our findings into a \textbf{temporal framework of newcomer persistence} (Table \ref{tab:persistence_framework}): (1) Acclimatization, (2) Acculturation, and (3) Embedding. For each stage, we identify the critical barriers to participation, the coping strategies used by survivors, and the corresponding design interventions required to scaffold successful integration}. 

\subsection{\abrev{Stage 1: Acclimatization}}
\abrev{The first distinct layer of the newcomer journey involves negotiating the physical reality of the virtual body. Our findings show that newcomers encounter challenges in social VR that go beyond those documented in 2D online spaces~\cite{Morgan2018Evaluating, Steinmacher2015Social}. Specifically, successful integration requires users to adapt to two immediate friction points: VR sickness and complex interfaces.}

\abrev{VR sickness (Sec 4.1.3) emerged as a particularly serious issue.} Unlike prior work that attributes sickness mainly to limited exposure or hardware limitations~\cite{hill2000habituation, howarth2008characteristics}, our analysis suggests the social platform design exacerbates these biological realities. Newcomers often stumble into high-intensity worlds without warning because the platform lacks adequate ``comfort ratings'' or sensory calibration during onboarding. \abrev{Consequently, in the absence of systemic support, the burden of physiological regulation shifts to the user. Our analysis of persistent users suggests that overcoming this barrier requires more than just building physical tolerance; it requires active management. Successful newcomers developed specific habits to navigate discomfort, such as identifying "safe" static worlds and strictly limiting session times (Sec. 4.1.3).}

\abrev{We argue that this self-regulation is necessitated by the high social cost of physiological failure, highlights a critical distinction between Social VR and single-player VR experiences. In a single-player game, physiological discomfort results in a pause with no social consequence. In Social VR, an inability to manage sickness forces an involuntary exit from social interactions. Therefore, physical comfort becomes a non-negotiable prerequisite for social presence. As established in presence theory, deep immersion relies on the user forgetting the medium~\cite{slater1997framework} and nausea draws immediate attention back to the physical body, breaking the social connection. Consequently, onboarding must do more than teach controls. It must scaffold users in identifying their physiological boundaries so they can remain present in the conversation without reaching a breaking point.}

\abrev{Furthermore, we argue that technical struggles also carry a higher social cost in collaborative environments.} While participants generally found basic interaction methods intuitive, they struggled with novel functions unique to 3D environments (Sec. 4.1.4), often overlooking features like spatial audio controls. \abrev{In a single-player context, a user struggling with a controller can troubleshoot in isolation. In Social VR, the interaction continues in real-time. Our findings show that a failure to manage the interface (e.g., failing to unmute or accidentally blocking someone) often results in a social error, not just a technical one.} This extends previous studies on VR onboarding, which have largely emphasized operational training (e.g., controller use) in offline or laboratory settings~\cite{bozgeyikli2017effects, chauvergne2023user, thoravi2020transceivr}. Such controlled exposures~\cite{wu2020exploring, Thoravi14} fail to reflect the urgency of the wild, \abrev{where newcomers must solve technical problems while under the pressure of observation. Thus, successful newcomers were distinctively those who learned to troubleshoot interface issues without letting them disrupt their social presence.}

\subsection{\abrev{Stage 2: Acculturation}}
\abrev{Once newcomers negotiate the physical reality of VR, they encounter a unique psychological challenge on the pressure of embodied social presence.}
Our findings indicate that social interaction in VR is fundamentally different from traditional online communication. Participants explicitly described 2D voice chats as feeling like ``talking on the phone'' (P15), allowing them to speak freely without pressure. In contrast, VR interactions induced the ``visceral anxiety of real-life socializing'' (Sec. 4.1.5).

\abrev{Viewing this through the LPP framework \cite{lave1991situated}, we identify a critical friction: the nature of the periphery itself. In 2D platforms, newcomers can lurk invisibly to observe norms \cite{preece2004top}. In Social VR, however, the periphery is embodied.  Newcomers are not just learning a platform interface, instead they are attempting to demonstrate the shared competence of embodied socialization required to enter the constellation of communities. Because their avatars are visible the moment they enter a room, they cannot observe without being observed. Consequently, the primary challenge is not just acquiring skills, but navigating this \textit{visible periphery} without the safety net of invisibility.}

This learning process is compounded by the ambiguity of social rules.
Adapting to social norms (Sec. 4.1.1) is fraught with uncertainty: behaviors intended as playful by veterans (e.g., physical touches) could be perceived as uncomfortable and problematic by newcomers. Prior research has documented harassment as a widespread and severe issue in social VR \cite{wei2025systematic,blackwell2019harassment}.
Yet, our participants rarely described their experiences using terms like ``harassment'' or ``toxicity.'' Instead, they framed their discomfort as part of the challenge of learning local norms.
This does not negate the reality of the toxic behavior; rather, it highlights that entry into these communities requires newcomers to decipher a ``shared repertoire'' of behaviors where the boundary between ordinary interaction and harmful conduct is not fixed but negotiated.
Survivors are distinctively those who look to veterans to distinguish between valid practice and abuse, reframing this ambiguity as a learning curve.
In this sense, the norms themselves entangle sociability with elements of toxicity, requiring newcomers to be especially attuned to the inappropriateness of specific interactions to gain legitimacy.

\abrev{The necessity of this negotiation highlights a gap in current platform design.}
Our findings extend prior work on newcomer integration that has primarily focused on 2D platforms \cite{Morgan2018Evaluating, Steinmacher2015Social} or small, homogeneous laboratory contexts \cite{wu2020exploring, Thoravi14}. Such controlled exposures fail to reflect the diversity of behaviors in the wild, where the lack of access to masters creates both opportunities and risks.
\abrev{Without targeted support to facilitate legitimate peripheral participation, many users likely disengage before building commitment.
Whether through overt harassment or the passive rejection of being ignored (exclusivity), the lack of normative guidance acts as a powerful filter. Therefore, onboarding must go beyond operational training. It must scaffold users in decoding social ambiguity and preparing them to recognize and respond to harassment, ensuring that the freedom of the open world does not become a barrier to entry.}

\subsection{\abrev{Stage 3: Embedding}} 
\abrev{If a newcomer survives the physiological challenges and navigates the social ambiguity, they need to find a reason to stay.} Open-world platforms like VRChat, and the envisioned Metaverse, are often defined by their lack of linear narrative, emphasizing user freedom and diverse social experiences~\cite{sykownik2021most,A2023METAVERSE}. However, our findings show that this very flexibility produces integration barriers for newcomers. Unlike traditional games that guide users with clear quests or objectives, Social VR offers little structural guidance on how to participate. New users frequently described feelings of purposelessness (Sec. 4.1.2), struggling to identify compelling activities amidst the vast amount of user-generated content. \abrev{Consequently, the primary challenge at this stage is shifting from shifting from the safety of the periphery (passive observer) to the center (active participant with a clear role).}

\abrev{This challenge is exacerbated by the platform's reliance on self-directed exploration rather than algorithmic curation. We argue that unlike 2D social media that sustain engagement through automated content delivery~\cite{Abi2023MATHEMATICAL,Qin2022The}, Social VR imposes significant agency demands on the newcomer. Content is not pushed to the user; it must be actively located. In the absence of an automated activity stream, newcomers must physically navigate between virtual worlds and initiate interactions to derive value from the platform.}

\abrev{Our analysis suggests that persistent users are characterized by their ability to transition from passive consumption to proactive discovery. These users succeed by generating their own intrinsic goals, such as systematic world-exploration or community seeking, to compensate for the lack of extrinsic structural guidance. Therefore, successful integration relies on establishing a social connection to replace the missing structural guidance. However, while exploration provides initial novelty, long-term embedding relies on establishing a social connection.} In the absence of game mechanics, social interaction becomes the primary metric of progression.  Survivors often described a pivotal moment, such as meeting a mentor or becoming a regular in a specific room, that transformed the empty open world into a familiar community. Without this social anchor, the environment remains a lonely expanse. 

\abrev{This points to a broader design implication: reliance on user freedom without scaffolding creates inequities in participation. Current designs rely heavily on serendipity. To support broader adoption, platforms should move beyond pure freedom and design intentional bridging mechanisms, such as structured newcomer events or mentorship matching, to help users find their purpose before they disengage.}

\subsection{Design Implications to Scaffold Newcomers Persistence} 
\abrev{Building on our persistence framework, we outline design implications structured by the three stages of newcomer persistence. Our core argument is that platforms must shift the burden of integration from the user to the system.}

\subsubsection{\abrev{Stage 1 Support: Facilitating Acclimatization}} 
\abrev{To support negotiating the physical reality of the virtual body and the complexity of the interface, platforms must address both physiological risks and technical unfamiliarity through systemic regulation.}

\textbf{\abrev{Systemic Safety.}} VR sickness poses a significant barrier to new user retention. This issue is exacerbated by newcomers' lack of experience with VR environments.
 To mitigate this challenge, one effective approach can be to begin with stationary experiences and slow-paced movement, allowing users to adjust to the immersive environment in a controlled manner. For example, a system may recommend (or only allow) users to explore a virtual room using \textit{teleportation}\footnote{A navigation method in VR, where users \textit{instantly} move to a selected point in the virtual environment. It can avoid motion sickness compared to locomotion, which is a common continuous movement technique where users traverse the virtual space, often using joysticks, walking in place, simulating real-world travel \cite{clifton2020effects}.}, supported by minimal distractions and stable reference points to help them build familiarity with VR mechanics while minimizing discomfort.
Default settings also play a crucial role in reducing VR sickness. Systems should automatically enable ``comfort-first'' options, such as slower movement speeds \cite{chang2020virtual}, and a highly reduced field of view during motion \cite{Fernandes2016}, gradually unlocking advanced settings as users gain confidence. Environmental design can further support comfort. Beginner-friendly spaces should feature even terrain, consistent visuals, and static horizon lines or objects to minimize disorientation. Moreover, virtual environments could include \textit{sickness risk level} labels, such as \textit{Comfortable}, \textit{Intermediate}, or \textit{Challenging}, to help newcomers choose spaces that match their current tolerance, further reducing the risk of discomfort and early negative experiences. \abrev{This allows newcomers to self-select environments that match their current tolerance, preventing the accidental exposure to high-intensity worlds described in Section 5.1.}

\textbf{\abrev{Scaffolding Interactions.}} \abrev{To assist newcomers in moving from familiar 2D screens to complex VR interactions, we propose a gradual interface progression. Platforms should not force newcomers to immediately master complex spatial inputs. Instead, essential functions (e.g., Mute, Block) should initially rely on familiar 2D paradigms (e.g., tablet-style menus) to reduce cognitive load. Crucially, access to these tools should be context-aware rather than front-loaded: for example, if a user enters a noisy instance, the system should dynamically highlight the "Audio Shielding" feature, ensuring survival tools are discovered via relevance rather than abstract tutorials.}

\abrev{As users graduate to unique VR interactions, such as gesture-based shortcuts, text instructions are insufficient. Drawing on recent findings in bare-hand guidance~\cite{wang2024exploring}, platforms should utilize ``Ghost Visualizations'', i.e., dynamic overlays that superimpose a semi-transparent hand over the user's avatar. Unlike static tooltips, these visualizations provide real-time error and target feedback, allowing newcomers to learn complex VR mechanics through embodied mimicry rather than trial-and-error.}

\subsubsection{\abrev{Stage 2 Support: Scaffolding Acculturation}}
\abrev{Design must reduce ambiguity and lower the stakes of early social interaction to support navigating social norms and managing safety.}

\textbf{\abrev{Visualizing the Hidden Curriculum.}} 
\abrev{Newcomers often hesitate to label discomfort as harassment, interpreting ambiguous aggression as local culture. To decouple safety from social anxiety, platforms should integrate normative calibration into onboarding, akin to pre-flight safety briefings. Rather than static text, this could involve situational simulations that distinguish between acceptable roughhousing and toxic behavior} (e.g., an avatar encroaching on personal space despite warnings). \abrev{By framing safety tools (e.g., ``Block/Report'') as standard operating procedure rather than nuclear options, platforms signal that using them is a valid, expected response to discomfort. This pre-validation empowers newcomers to trust their instincts, reducing the cognitive load of evaluating harassment in real-time.}

\textbf{\abrev{Architecting Legitimate Peripheral Participation.}} 
\abrev{To enable newcomers to observe social dynamics without the immediate threat of performance, platforms must facilitate LPP through spatial design~\cite{lave1991situated}. In 2D platforms, users can lurk invisibly; in Social VR, the avatar implies presence and availability. To resolve this, we propose the design of ``Spatial Buffers'': observation decks, balconies, or foyer areas where newcomers spawn by default. These zones allow users to read the room and decipher local norms from a safe distance, visually separated from the main social floor. This spatial affordance legitimizes peripheral participation, reframing passive observation from an ambiguous social signal into a structured stage of acculturation. It enables newcomers to modulate their social presence, transitioning from the periphery to the center only after establishing normative comfort.}

 \subsubsection{\abrev{Stage 3 Support: Fostering Embedding}}
\abrev{The final stage requires helping newcomers find purpose and community within the open-ended environment. Design must shift from purely spatial solutions to social and structural scaffolding.}

\textbf{\abrev{Reducing Agency Demands via Intelligent Scaffolding.}} 
\abrev{Our findings highlight that the lack of structured paths leads to disorientation (Sec. 4.1.2). While human guides are effective, they are not scalable. To bridge this gap, we propose the integration of \textit{AI-based Embodied Agents} to act as personalized, scalable mentors. Drawing on recent success in MMORPGs, where LLM-based agents like AMAN have effectively guided novices through complex mechanics~\cite{lee2025aman}, Social VR platforms could deploy similar agents to provide context-aware orientation.} 
\abrev{Unlike static tutorials, these agents could respond dynamically to user behavior—answering questions like ``Is it normal to do X here?'' or ``Where can I find the aviation community?''—thereby reducing the social ambiguity that paralyzes newcomers. By offloading the "instructional" labor to agents, human interactions can focus on social bonding rather than technical troubleshooting.}

\textbf{\abrev{Gamifying Exploration with Community Passports.}} 
\abrev{To further mitigate purposelessness, platforms can implement \textit{Community Passports}: lightweight, gamified objectives that reward newcomers for visiting specific hubs or attending events. Unlike rigid quests, these passports act as a structural excuse to explore, giving newcomers a valid reason to enter social spaces without the awkwardness of having no clear agenda. By formalizing exploration (e.g., ``Visit 3 Music Worlds''), the platform creates a legitimate periphery, allowing users to linger in social spaces under the guise of completing a task while observing local norms.}

\textbf{Fostering Inclusivity with Experienced Users.}
Veterans forming insular groups can exacerbate the integration challenges for newcomers, who may find these groups unwelcoming or difficult to penetrate.
Design and policy initiatives that actively promote inclusivity in community interactions are essential for creating a supportive environment that eases the integration of new users. One effective approach can be establishing mentorship programs where experienced users guide newcomers through both the social and technical aspects of the platform. Such programs could pair veterans with newcomers, encouraging interaction and knowledge-sharing while reducing the barriers to entry for less experienced users. Mentors might assist with navigation, introduce platform features, and facilitate participation in community events, helping newcomers feel more confident and engaged. Incentivizing participation through rewards for veteran users, such as exclusive badges, in-platform recognition, or access to special features, could further encourage active mentorship and foster a cooperative community atmosphere. \abrev{This transforms veteran exclusivity into a structured role, providing the newcomer with the social anchor necessary for long-term retention.}

\section{Limitations and Future Work} 
Our study does not include individuals who tried VRChat only once or twice before leaving, for methodological reasons discussed earlier. \abrev{We acknowledge that our sample, consisting of users who successfully integrated into VRChat, likely possesses higher baseline digital literacy, gaming experience, or social resilience compared to users who quit immediately. Consequently, our findings may not capture the specific breaking points of less tech-savvy or more introverted populations.} \abrev{To uncover the specific aspects of social VR that are immediately discouraging or intolerable, future research should explicitly target early quitters and map the lower bounds of tolerance that our persistent participants successfully exceeded.}
Another direction is to explore how harassment is entangled with newcomer integration. Participants in our study described discomfort or ambiguous encounters, but harassment itself was not a central focus. Yet harassment is a well-documented issue in social VR. 
Future research could therefore employ more focused approaches to explore how it intersects with newcomer integration. Such work would benefit from considering both perspectives: newcomers’ accounts can illuminate how boundary-crossing behaviors affect their early experiences, while established users’ accounts can show how newcomers’ presence may disrupt or challenge existing norms. Considering these perspectives together would provide a more balanced understanding of how harassment and integration are connected, and how onboarding and community design might reduce risks while supporting inclusive participation.
\abrev{Finally, as Social VR demographics expand \cite{fiani2024exploring}, future work must also consider the perspectives of younger users (children and teenagers) to understand how age and prior gaming experience moderate the severity of onboarding challenges and the effectiveness of persistence strategies.}

\section{Conclusion}
 Our research contributes to the understanding of newcomer integration in online VR platforms. We identified unique challenges faced by newcomers, such as adapting to novel user interfaces, comprehending new social norms, and managing sensory overload, which extend traditional models of online community engagement to the immersive VR context. Our findings reveal the necessity of addressing not just operational aspects like controller use, but also issues like VR sickness in onboarding processes. Furthermore, we emphasize the difficulties posed by the lack of clear objectives in social VR environments and how this affects user retention and satisfaction. Additionally, we provide design strategies to enhance inclusivity and navigability, facilitating the integration and retention of newcomers in VR spaces.

\section*{\abrev{Use of Generative AI Tools}}
\abrev{A generative AI tool was used to improve the grammar and clarity of the authors’ original text. It did not contribute to scientific content including to generate novel ideas, data analysis, or interpretation. All scientific work was performed and validated by the authors.}

\bibliographystyle{ACM-Reference-Format}
\bibliography{sample-base}

\clearpage
\appendix
\label{app:myappendix}

\section{Appendix}

\begin{table}[h!]
\centering
\begin{tblr}{
  width = \linewidth,
  colspec = {Q[117]Q[50]Q[112]Q[131]Q[181]Q[335]},
  cells = {c},
  hlines,
  hline{1,26} = {-}{0.08em},
}
\textbf{Participants} & \textbf{Age} & \textbf{Gender} & {\textbf{Status}~} & \textbf{Experience} & \textbf{Used Platforms}~\\
P1 & 22 & Male & New & Less than 1 month~ & VRChat\\
P2 & 27 & Female & Experienced & ~ 3.5 years~ & VRChat, RecRoom\\
P3 & 20 & Female & New & 3 months~ & VRChat, Horizon Worlds\\
P4 & 25 & Male & New & Less than 1 month~ & VRChat\\
P5 & 24 & Male & Experienced & 6.5 years & VRChat, RecRoom, AltspaceVR\\
P6 & 22 & Male & New & 2 months~ & VRChat\\
P7 & 26 & Female & Experienced & 1.5~ years & VRChat, RecRoom\\
P8 & 29 & Male & Experienced & 8 months & VRChat, RecRoom\\
P9 & 24 & Male & New & Less than 1 month~ & VRChat\\
P10 & 26 & Female & Experienced & 1 year & VRChat, RecRoom~\\
P11 & 18 & Female & New & 2 months~ & VRChat\\
P12 & 25 & Male & New & Less than 1 month~ & VRChat, RecRoom\\
P13 & 24 & Non-binary & New & Less than 1 month~ & VRChat\\
P14 & 20 & Male & Experienced & 2 years & VRChat, RecRoom\\
P15 & 33 & Male & New & Less than 1 month~ & VRChat\\
P16 & 24 & Female & New & 1 Month & VRChat\\
P17 & 18 & Male & New & 2 months~ & VRChat\\
P18 & 26 & Female & Experienced & 4 years & VRChat, Horizon Worlds, AltspaceVR\\
P19 & 27 & Male & Experienced & 1.5 year & VRChat\\
P20 & 33 & Male & Experienced & 10 Months & VRChat, Sansar\\
P21 & 21 & N.A. & Experienced & 3 years & VRChat, ChilloutVR, AltspaceVR\\
P22 & 29 & Male & New & 2 months & VRChat\\
P23 & 19 & Female & Experienced & 1 year & VRChat, RecRoom\\
P24 & 21 & Male & New & Less than 1 month~ & VRChat
\end{tblr}
\caption{\textbf{Information of Participants}}
\label{tab:interviewee}
\end{table}

\end{document}